\documentstyle[preprint,aps,eqsecnum]{revtex}

\def\beq#1{\begin{equation} \label{#1}}
\def\eeq{\end{equation}}
\newcommand{\bea}{\begin{eqnarray}}
\newcommand{\eea}{\end{eqnarray}}
\def\bra#1{\left\langle #1\right\vert}
\def\ket#1{\left\vert #1\right\rangle}

\def\PLB{{ Phys. Lett.} B}

\def\PRD{{ Phys. Rev.} D}
\begin{document}
{
\tighten
  
\title {Stodolsky's Theorem and Neutrino Oscillation Phases  -- for pedestrians}
\author{Harry J. Lipkin\,\thanks{Supported
in part by grant from US-Israel Bi-National Science Foundation
and by the U.S. Department
of Energy, Division of High Energy Physics, Contract W-31-109-ENG-38.}}
\address{ \vbox{\vskip 0.truecm}
  Department of Particle Physics
  Weizmann Institute of Science, Rehovot 76100, Israel \\
\vbox{\vskip 0.truecm}
School of Physics and Astronomy,
Raymond and Beverly Sackler Faculty of Exact Sciences,
Tel Aviv University, Tel Aviv, Israel  \\
\vbox{\vskip 0.truecm}
High Energy Physics Division, Argonne National Laboratory,
Argonne, IL 60439-4815, USA\\
~\\lipkin@hep.anl.gov
\\~\\
}
 
\maketitle
 
\begin{abstract} 
Neutrino oscillations are experimentally observable only as a result of
interference between neutrino states with different masses and THE SAME
ENERGY. All interference effects between neutrino states having different
energies are destroyed by the intereaction between the incident neutrino 
and the neutrino detector. Erroneous results are frequently obtained by
neglecting the neutrino-detector interactions. 
\end{abstract} 
} 

Stodolsky\cite{Leo} has given a very simple answer to the confusion that
still arises in discussions of the phase of neutrino oscillations.  The
relevant literature producing this confusion has recently been summarized
and clarified\cite{Okun}.  The purpose of this note is to support this
excellent analysis\cite{Okun} without engaging in a direct debate against
the confusing articles and also to present a ``pedestrian" version of
Stodolsky's work which is hopefully understandable to students and
experimentalists.
  
The detection of a neutrino always involves its interaction with a
detector that is part of an environment described by a density matrix in
which the energy is diagonal. Unless this interaction with the environment
is turned off, and no experiment can do this, all relative phase
information between neutrino states with different energies is destroyed.

      The question is not whether states of the same momentum and different
energies are coherent, states of the same energy  and different momentum or
states of the same velocity. There have been many irrelevant arguments about
these issues. But states with different ENERGIES ARE NEVER COHERENT in any
realistic experiment. States of the same energy and different momenta can be
coherent, but may not be. This depends upon the way the measurement is made.
But states with different energies can not be coherent.

This discussion refers only to neutrino detectors,   The usual detector is a
nucleon, which changes its state after absorbing a neutrino and emitting a
charged lepton, and is initially either in an energy eigenstate or in a
statistical mixture in thermal equilibriam with its surroundings.  
No neutrino detector has ever been prepared in a coherent mixture of energy
eigenstates and no such detector has been proposed for future experiments .

      All arguments about Lorentz invariance are irrelevant. The
detector chooses a particular Lorentz frame where the detector is at rest and
described by a density matrix in which the energy is diagonal and no
interference between states of different energies can be observed.

 Most treatments  do not consider at all the quantum mechanics of the detector.
Since the detector is a quantum system (e.g. a nucleon) which undergoes a
transition together with the neutrino-to-charged-lepton transition, and the
initial and final states of the detector are not measured, the transition
probability is the square of the transition matrix element for the whole
system, averaged over detector initial states and summed over detector final
states. This immediately kills all interference between neutrino states with
different energies as they are accompanied by different detector states which
must have different energies  because  energy in conserved in the process. The
detector states with different energies  are orthogonal to one another and all
interference terms between them vanish because of this detector orthogonality.

In this context the ``factor-of-two" arguments are seen to be 
missing an essential point in the actual neutrino oscillation experiments;
namely the role of the detector as a quantum-mechanical system entangled  with
the neutrino.

The standard textbook neutrino-oscillation wave function, a coherent
linear combination of states with different energies, is not found in real
experiments. Thus considerable confusion remains even though coherence,
interference and dephasing have been extensively discussed and clarified
\cite{Leo,Dost,QM,NeutHJL,Kayser,GoldS,Pnonexp,MMNIETO,GrossLip,okun1,pnow98}.
Elementary quantum mechanics and quantum statistical mechanics tell us
that the components of the density matrix describing a neutrino detector
and having different energies are never coherent\cite{Leo}, while neutrino
components with different masses and different momenta must be coherent to
cancel components with the wrong flavor just outside the neutrino source..
This coherence between source states having the same energy and different
momenta can produce coherence between neutrino states with the same energy
and different masses.

This physics is illustrated in detail in a  toy model\cite{pwhichfin} for
the detection of a neutrino as a
transition between an initial state of a neutrino and a detector and a
final state of a muon and the same detector.  The wave function for the 
initial state of neutrino and detector is
\beq{WP21}
\Psi_i(\nu,D) = 
\sum_{k=1}^{N_\nu} \ket{\nu(E_\nu,m_k,\vec P_k),D_i(E_i)}  
 \eeq
where $N_\nu$ is the number of neutrino mass states, 
$E_\nu$, $m_k$ and $\vec P_k$ denote the neutrino energy, mass and momentum and 
$D_i(E_i)$ is
the initial state of the detector with energy 
$E_i$. 
If the detector is a muon detector the final detector state after neutrino 
absorption is 
\beq{WP23}
 \Psi_f(\mu^\pm,D) = 
\sum_{k=1}^{N_\nu} \ket{\mu^\pm(E_\mu,\vec P_\mu),D^\mp_{kf}(E - E_\mu)}  
 \eeq
where  
$E_\mu$ and $\vec P_\mu$ denote the muon energy and momentum,  $D^\mp_{kf}$ is
the final detector state produced in the ``path $k$"; i.e. after the absorption
of a neutrino with mass  $m_k$ and emission of a $\mu^\pm$, and  
$E=E_\nu + E_i$ is the total energy which is conserved in the transition. 

The transition in the detector occurs on a nucleon, whose co-ordinate is
denoted by by $\vec X$, and involves a charge exchange denoted by the isospin
operator $I_{\mp}$ and a momentum    transfer $\vec P_k -\vec P_\mu$.
The detector transition matrix element is therefore given by 
\beq{WP24} 
\bra {D^\mp_{kf}} T^{\mp}\ket {D_i} =  
\bra {D^\mp_{kf}}I_{\mp}e^{i(\vec P_k -\vec P_\mu) \cdot \vec X}  \ket {D_i} 
 \eeq 
 
The overlap between the final detector wave functions after the transitions 
absorbing neutrinos with masses $m_k$ and $m_j$ is then 
 \beq{WP25} 
\langle {D^\mp_{kf}}\ket {D^\mp_{jf}} =  
 \bra {D_i} e^{i(\vec P_j-\vec P_k) \cdot \vec X}  \ket{D_i}  \eeq

If the quantum fluctuations in the position of the active nucleon in the
initial state of the detector are small in comparison with the oscillation wave 
length, $\hbar /(\vec P_j-\vec P_k)$,
\beq{WP26}
|\vec P_j -\vec P_k|^2 \cdot \bra {D_i} |\vec X^2|\ket {D_i} \ll  1      
 \eeq 
\beq{WP27}
\langle {D^\mp_{kf}} \ket {D^\mp_{jf}} \approx  
1 - (1/2)\cdot|\vec P_j -\vec P_k|^2 \cdot \bra {D_i} |\vec X^2|\ket {D_i} 
\approx 1
 \eeq

There is thus effectively a full overlap between the final detector states after
absorption of different mass neutrinos, and a full coherence between the neutrino
states with the same energy and different momenta. 

The total energies of the final muon and detector produced after absorption of
neutrinos with different energies are different. These muon-detector states are thus
orthogonal to one another and there is no coherence between detector states produced
by the absorption of neutrinos with different energies. 

There have been suggestions for bypassing Stodolsky's theorem by
exploiting some kind of energy-time uncertainty to detect interference
between components having different energies in the neutrino wave
function. The time of flight of the neutrino from source to detector might
be measured by detecting the muon emitted together with the neutrino in a
pion decay in the source and measuring precisely the times of emission in
the source and of absorbtion in the detector.

However, if the quantum fluctuations in the position of the active nucleon
in the initial state of the detector are small in comparison with the
oscillation wave length, eqs. (\ref{WP26}) and (\ref{WP27}) apply and the
coherence and relative phase of the components in the neutrino wave
function having the same energy and different momenta are preserved . This
relative phase completely determines the flavor output of the detector;
i.e. the relative probabilities of producing a muon or an electron. These
probabilities in all realistic cases are essentially independent of energy
over the relevant energy range. Thus the relative phases and coherence
between components in the neutrino wave function with different energies
is irrelevant. All energies give the same muon/electron ratio whether they
add coherently or incoherently and time measurements cannot change the
muon/electron ratio observed at the detector.

It is a pleasure to thank Maury Goodman, Yuval Grossman,  Boris Kayser, Lev
Okun,and Leo Stodolsky for helpful discussions and comments.


\begin{references}
\bibitem{Leo}{Leo Stodolsky, Phys. Rev. D58 (1998) 036006}  

\bibitem{Okun} {L. B. Okun et al.  hep-ph/0211241}

\bibitem{Dost}{H.J. Lipkin, Neutrino Oscillations for Chemists,
Lecture Notes (unpublished).}
 
\bibitem{QM}{H.J. Lipkin, Lecture Notes on Neutrino Oscillations for a Quantum
Mechanics course (unpublished).}
 
\bibitem{NeutHJL}{H.J. Lipkin, Neutrino Oscillations and MSW
for Pedestrians, Lecture Notes (unpublished).}
 
\bibitem{Kayser}{B. Kayser, \PRD\,24 (1981) 110.}
 
\bibitem{GoldS}
{T. Goldman, hep-ph/9604357.}
 
\bibitem{Pnonexp}
{H.J. Lipkin, \PLB\,348 (1995) 604.}
 
\bibitem{MMNIETO}
{M.M. Nieto, hep-ph/9509370, Hyperfine Interactions 100 (1996) 193}
 
\bibitem{GrossLip} {Yuval Grossman and Harry J. Lipkin,
Phys. Rev. D55 (1997) 2760}
 

\bibitem {okun1}{   A.D. Dolgov, A.Yu. Morozov, L.B. Okun, and M.G. Schepkink,
{\it Nucl. Phys.} {\bf 502} (1997) 3}

\bibitem{pnow98}{ Harry J. Lipkin, hep-ph/9901399, Weizmann Preprint
WIS-98/31/Dec-DPP, Tel Aviv University preprint TAUP 2537-98,  Argonne Preprint
ANL-HEP-CP-98-126, available in Proceedings of the Europhysics Neutrino 
Oscillation Workshop (NOW'98) 7-9 September 1998. Amsterdam. Published in
http://www.nikhef.nl/pub/conferences/now98/}

\bibitem{pwhichfin}
{H.J. Lipkin, hep-ph/9907551, \PLB\,477 (2000) 195}

\end{references}
\end{document}